\documentclass[12pt]{article}
\usepackage{epsfig}
\usepackage{float}
\usepackage{amssymb,amsmath}
\newcommand{\be}{\begin{equation}}
\newcommand{\ee}{\end{equation}}
\newcommand{\bea}{\begin{eqnarray}}
\newcommand{\eea}{\end{eqnarray}}
\begin{document}

\begin{center}
{\bf TIME-ENERGY UNCERTAINTY RELATIONS FOR  NEUTRINO OSCILLATIONS AND THE
M\"OSSBAUER NEUTRINO EXPERIMENT}
\end{center}
\begin{center}
S. M. Bilenky
\end{center}

\begin{center}
{\em  Joint Institute for Nuclear Research, Dubna, R-141980,
Russia\\}
\end{center}
\begin{center}
F. von  Feilitzsch and W. Potzel
\end{center}
\begin{center}
{\em Physik-Department E15, Technische Universit\"at M\"unchen,
D-85748 Garching, Germany}
\end{center}

\begin{abstract}
Using the Mandelstam-Tamm method we derive time-energy uncertainty
relations for neutrino oscillations. We demonstrate that the small
energy uncertainty of  antineutrinos in a recently considered
experiment with recoilless resonant (M\"ossbauer) production and
absorption of tritium antineutrinos is in conflict with the energy
uncertainty which, according to the time-energy uncertainty
relation, is necessary for neutrino oscillations to happen. Oscillations of M\"ossbauer neutrinos would indicate a stationary phenomenon where the evolution of the neutrino state occurs in space rather than in time. A
M\"ossbauer neutrino experiment could provide a unique possibility
to reveal the true nature of neutrino oscillations.
\end{abstract}

\section{Introduction}
The observation of neutrino oscillations in the Super-Kamio\-kande
atmos\-pheric \cite{SK}, SNO solar \cite{SNO}, KamLAND reactor
\cite{Kamland} and other neutrino experiments
\cite{Cl,Gallex,Sage,SKsol,K2K,Minos} is one of the most important
recent discoveries in particle physics. Small neutrino masses can
not naturally be explained by the Standard Higgs mechanism. Their
explanation requires a new mechanism of neutrino mass generation
beyond the Standard Model. At present, the see-saw mechanism
\cite{seesaw}, which is based on the assumption of a violation of
the total lepton number at a scale which is much larger than the
electroweak scale, is considered as the most plausible mechanism of
neutrino mass generation. The see-saw mechanism requires for
neutrinos with definite masses  to be Majorana particles. The
discovery of the neutrino-less double $\beta$-decay (see
\cite{Fiorini}), which is allowed only if massive neutrinos are
Majorana particles, would be a strong evidence in favor of the
see-saw idea.

Existing neutrino oscillation data are perfectly described if we
assume the three-neutrino mixing (see \cite{BGG,Conca})
\begin{equation}\label{mix}
\nu_{lL}(x)=\sum_{i=1}^{3}U_{li}\,\nu_{iL}(x)\quad (l=e,\mu,\tau)~.
\end{equation}
 Here $\nu_{lL}(x)$ is the field of the flavor neutrino
$\nu_{l}$, $\nu_{i}(x)$ is the field of neutrinos with mass $m_{i}$
and $U$ is the $3\times3$ unitary PMNS \cite{Pont57,MNS} mixing
matrix which is characterized by three mixing angles $\theta_{12}$,
$\theta_{23}$ and $\theta_{13}$ and the $CP$ phase $\delta$.

The standard expression for the probability of the transition
$\nu_{l}\to \nu_{l'}$ has the form
\begin{equation}\label{transitionprob}
{\mathrm P}_{\nu_{l} \to \nu_{l'}} =|\sum^{3}_{i=1}U_{l' i} \,~ e^{-
i \Delta m^2_{2i} \frac {L} {2E}} \,~U_{l i}^*\, |^2~.
\end{equation}
Here $E$ is the neutrino energy, $L$ is the distance between the
neutrino source and the neutrino detector and $\Delta m^2_{ki}= m^2_{i}-
m^2_{k}$.

The probability ${\mathrm P}(\nu_{l} \to \nu_{l'})$ depends on six
parameters (two mass-squared differences, three angles and one
phase). From the analysis of the data of neutrino oscillation
experiments follows, however, that the parameter $\Delta
m^2_{12}$ is much smaller than $\Delta m^2_{23}$:
\begin{equation}\label{hierarchy}
\Delta m^2_{12}\simeq 3\cdot 10^{-2}~ \Delta m^2_{23}~.
\end{equation}
Thus,  at $\Delta m^2_{23} \frac {L} {2E}\gtrsim 1$, i.e., in the
atmospheric and accelerator long-baseline (LBL) region, in first
approximation we can neglect the small contribution of $\Delta
m^2_{12}$ to the transition probabilities. In this case, the
probability of $\nu_{l}$ to survive takes the two-neutrino form (see
\cite{BGG})
\begin{equation}\label{survival1}
{\mathrm P}_{\nu_{l} \to \nu_{l}}={\mathrm P}_{\bar\nu_{l} \to
\bar\nu_{l}}= 1-\frac {1} {2}~B_{ll}~(1-\cos\Delta m^2_{23} \frac
{L} {2E})~.
\end{equation}
Here
\begin{equation}\label{amplitude1}
B_{ll}=4 |U_{l3}|^{2}~(1-|U_{l3}|^{2})
\end{equation}
is the amplitude of the oscillations.

From (\ref{survival1}) we find the following expression for the
$\bar\nu_{e} \to \bar\nu_{e}$ survival probability
\begin{equation}\label{baresurvival}
{\mathrm P}_{\bar\nu_{e} \to \bar\nu_{e}}= 1-\frac {1}
{2}~\sin^{2}2\theta_{13}~(1-\cos\Delta m^2_{23} \frac {L} {2E})~.
\end{equation}
In the reactor experiment CHOOZ \cite{Chooz} no indications for
neutrino oscillations driven by $\Delta m^2_{23}$ were found. From
the exclusion plot obtained from the data of this experiment, the
following upper bound was found for the parameter
$\sin^{2}\theta_{13}$
\begin{equation}\label{teta13}
\sin^{2}\theta_{13}\lesssim  5\cdot 10^{-2}~.
\end{equation}
Neglecting the small contribution of $\sin^{2}\theta_{13}$, we find from (\ref{survival1}) and
(\ref{amplitude1}) in the atmospheric-LBL region of $\frac
{L} {E}$ the following expression for the
probability of $\nu_{\mu}$ to survive:
\begin{equation}\label{atmsurvival}
{\mathrm P}_{\nu_{\mu} \to \nu_{\mu}}\simeq 1-\frac {1}
{2}~\sin^{2}2\theta_{23}~(1-\cos\Delta m^2_{23} \frac {L} {2E})~.
\end{equation}
In the KamLAND region ($\Delta m^2_{12} \frac {L} {2E}\gtrsim 1$)
the probability of the transition $\bar\nu_{e}\to \bar\nu_{e}$ is
given by (see \cite{Schramm,BGG})
\begin{equation}\label{KLsurvival}
{\mathrm P}_{\bar\nu_{e} \to \bar\nu_{e}}
=|U_{e3}|^{4}+(1-|U_{e3}|^{2})^{2}~\frac{1}{2}\sin^{2}2\theta_{12}~(1-\cos\Delta
m^2_{12} \frac {L} {2E})~.
\end{equation}
Thus, in the approximation $|U_{e3}|^{2}\to 0$ we have
\begin{equation}\label{twoneutrinoKL}
{\mathrm P}_{\bar\nu_{e} \to \bar\nu_{e}} \simeq 1-\frac {1}
{2}~\sin^{2}2\theta_{12}~(1-\cos\Delta m^2_{12} \frac {L} {2E})~.
\end{equation}
The expressions (\ref{atmsurvival}) and (\ref{twoneutrinoKL}) are
widely used in the analysis of the neutrino oscillation data. From the
analysis of the data of the atmospheric Super-Kamiokande experiment,
the following 90\% CL ranges were obtained for the parameters $\Delta m^{2}_{23}$
and $\sin^{2}2 \theta_{23}$ \cite{SK}
\begin{equation}\label{SKrange}
 1.9\cdot 10^{-3}\leq \Delta m^{2}_{23} \leq 3.1\cdot
10^{-3}\rm{eV}^{2},\quad \sin^{2}2 \theta_{23}> 0.9.
\end{equation}
The following best-fit values of the parameters were found in
\cite{SK}
\begin{equation}\label{SKbest}
\Delta m^{2}_{23}=2.5\cdot 10^{-3}\rm{eV}^{2},\quad \sin^{2}2
\theta_{23}=1.
\end{equation}
The results of the Super-Kamiokande experiment have perfectly been
confirmed by the accelerator K2K \cite{K2K} and MINOS \cite{Minos}
long baseline neutrino oscillations experiments. The analysis
of the MINOS data gave the result \cite{Minos}:
\begin{equation}\label{Minosdata}
\Delta m^{2}_{23}=(2.38^{+0.20}_{-0.16})\cdot
10^{-3}\rm{eV}^{2},\quad \sin^{2}2 \theta_{23}>0.84~(90\% ~CL).
\end{equation}
From the global analysis of the recent data of the reactor
experiment KamLAND and the data of the solar neutrino experiments
the following values were obtained for the parameters $\Delta
m^{2}_{12}$ and $\tan^{2} \theta_{12}$ \cite{Kamland}:
\begin{equation}\label{KLsolar}
\Delta m^{2}_{12} = (7.59^{+0.21}_{-0.21})\cdot
10^{-5}~\rm{eV}^{2},\quad\tan^{2} \theta_{12}= 0.47^{+0.06}_{-0.05}
\end{equation}

Concerning the study of neutrino oscillations,
a new stage of high-precision experiments starts at present.
In the future DOUBLE CHOOZ
\cite{Doublechooz} and Daya Bay \cite{Dayabay} reactor neutrino
experiments, the sensitivities to the parameter $\sin^{2}2 \theta_{13}$
will be about 10-20 times better than in the CHOOZ experiment. The
same sensitivity is planned to be reached in the accelerator T2K
experiment \cite{T2K}. In the latter, the parameters $\Delta
m^{2}_{23}$ and $ \sin^{2}2 \theta_{23}$ will be measured with the
accuracies $\delta(\Delta m^{2}_{23})\sim 10^{-4}\rm{eV}^{2}$ and
$\delta (\sin^{2}2 \theta_{23})\sim 10^{-2}$, respectively. High
precision neutrino oscillation experiments are planned at the future
Super Beam, Beta Beam, and Neutrino Factory facilities
\cite{NuFact07}.

In spite of the big progress in the investigation of neutrino
oscillations and of future prospects there was not so much progress in
the understanding of the physics  of neutrino oscillations.

Two factors of different origin determine the neutrino transition
probability (\ref{transitionprob}):
\begin{itemize}
  \item The elements of the mixing matrix $U$ which characterize
  the mixed states of flavor neutrinos
\begin{equation}\label{flavorstates}
    |\nu_{l}\rangle =\sum_{i}U^{*}_{li}|\nu_{i}\rangle~,
\end{equation}
where $|\nu_{i}\rangle$ is the state of a neutrino with the mass
$m_{i}$ and momentum $p_{i}$.
\item The oscillation phases
\begin{equation}\label{oscilphases}
    \phi_{ik}=\Delta m^2_{ik} \frac {L} {2E}
\end{equation}
which are determined by the evolution of the vectors
$|\nu_{i}\rangle$.
\end{itemize}
Different authors make different assumptions on the neutrino states
with definite masses $|\nu_{i}\rangle$ (same momenta and different
energies (see \cite{BilPont,Fritsch}), same energies and different
momenta (see \cite{Stodol,Lipkin,Kayser}), different momenta and
different energies (see \cite{Winter, BilGiunti,Giunti})) and
different assumptions on the evolution of these states (in time or
in space and time). These completely different physical assumptions
lead to the same standard expression (\ref{transitionprob}) for the
transition probability (see, for example, \cite{BilFeilPotz07}).
Thus, the study of neutrino oscillations in usual neutrino
oscillation experiments does not allow to distinguish different
hypotheses on the states of neutrinos with definite masses and their
evolution. We showed in \cite{BilFeilPotz07} that a new type of
neutrino experiment with resonant recoilless (M\"ossbauer) emission
and absorption of monochromatic $\bar\nu_{e}$ from atomic two-body
tritium decay, proposed in \cite{Raghavan, Potzel}, would allow to
discriminate different basic assumptions on the theory of neutrino
oscillations.

If neutrino oscillations are a non-stationary phenomenon, the
time-energy uncertainty relation holds for such a process (see
\cite{BilMat06,Bil07}). In the following section we will obtain the
time-energy uncertainty relation for neutrino oscillations using the
general Mandelstam-Tamm method \cite{TammMand45}. We will discuss
the M\"ossbauer neutrino experiment from the point of view of this
relation. We will show that the M\"ossbauer neutrino experiment
could allow to answer the fundamental question: is the time-energy
uncertainty relation applicable to neutrino oscillations, i.e.
are neutrino oscillations a non-stationary phenomenon?

\section{Time-energy uncertainty relation for neutrino oscillations}

Uncertainty relations play a fundamental role in quantum theory.
They are based on general properties of the theory and manifest its
nature. There are two different types of uncertainty
relations in quantum theory: the Heisenberg uncertainty relations
and the time-energy uncertainty relation.

All uncertainty relations are based on the Cauchy-Schwarz inequality
(see, for example, \cite{Messiah})
\begin{equation}\label{Couchy}
 \Delta A~\Delta B \geq \frac{1}{2}|~\langle
 \Psi|~[A,B]~|\Psi\rangle|~.
\end{equation}
Here $A$ and $B$ are Hermitean operators, $\Delta A$ and $\Delta B$
are the standard deviations  and $|\Psi\rangle$ is some state. We
have
\begin{equation}\label{standarddev}
 \Delta A=\sqrt{\langle \Psi|~(A- \bar A )^{2}~|\Psi \rangle} =
 \sqrt{\overline {A^{2}}- (\bar A )^{2}}~,
\end{equation}
where
\begin{equation}\label{average}
    \bar A=\langle
 \Psi|~A~|\Psi\rangle~.
\end{equation}
The Heisenberg uncertainty relations are a direct consequence of the
inequality (\ref{Couchy}) and the commutation relations for the
operators $A$ and $B$. For example, from the commutation relation
\begin{equation}\label{commutrelation}
[p,q]=\frac{1}{i}
\end{equation}
and the inequality (\ref{Couchy}), we derive the standard Heisenberg
uncertainty relation
\begin{equation}\label{Heisenuncertainty}
\Delta p~\Delta q \geq \frac{1}{2}
\end{equation}
for the operators of the momentum $p$ and the coordinate $q$.
Let us stress that the Heisenberg uncertainty relations for canonically
conjugated quantities have a universal character: the form of these
relations does not depend on the state $|\Psi\rangle$.

The time-energy uncertainty relation has a completely different
character. This is connected with the fact that time in quantum
theory is a parameter and  there is no operator which corresponds to
time.

The time-energy uncertainty relation is based on the fact that the
dynamics of a quantum system is determined by the Hamiltonian. The
most general method of derivation of the time-energy uncertainty
relation was given by Mandelstam and Tamm \cite{TammMand45}.

According to general principles of the quantum field theory,
in the Heisenberg representation for any
operator $O(t)$, which does not
depend upon time explicitly, we have (see, for example,
\cite{BogShirkov, Peskin})
\begin{equation}\label{evolution}
    i~\frac{ d ~O(t)}{d ~t}=[O(t),H]~,
\end{equation}
where $H$ is the total Hamiltonian (which does not depend on time).
From (\ref{Couchy}) and (\ref{evolution}) we find
\begin{equation}\label{timeenergy1}
    \Delta E~\Delta O(t) \geq \frac{1}{2}
   ~ |\frac{d }{d t} \overline{ O}(t)|~.
\end{equation}
Here
\begin{equation}\label{average1}
\overline{ O}(t)=\langle
 \Psi_{H}|~O(t)~|\Psi_{H}\rangle=\langle
 \Psi(t)|~O~|\Psi(t)\rangle~,
\end{equation}
where $O$ and $|\Psi(t)\rangle$ are the  operator and the vector of
the state in the Schr\"odinger representation.

It is evident from (\ref{timeenergy1}) that in the case of a
stationary state ($\Delta E=0$) for  any operator $O(t)$ the average
$\overline O(t)$ does not depend on $t$ .

The relation (\ref{timeenergy1}) can be written in the form of the
time-energy uncertainty relation
\begin{equation}\label{timeenergy2}
    \Delta E~\Delta t \geq \frac{1}{2}~,
\end{equation}
where
\begin{equation}\label{deltat}
\Delta t=\frac{\Delta O(t) }{| \frac{d }{d t} \overline{O}(t)|}~.
\end{equation}
The relation (\ref{timeenergy2}), unlike the Heisenberg uncertainty
relations, does not have a universal character. For example, for a
wave packet, $\Delta t $ is the time interval during which the wave
packet passes a fixed space point, for an excited state, $\Delta t $
is the life-time of the state, etc. (see\cite{TammMand45,Busch07}).

From the Mandelstam-Tamm relation (\ref{timeenergy1}) we will now
obtain the time-energy uncertainty relations for neutrino
oscillations. Let us choose $O= P_{l}$, where
\begin{equation}\label{projection}
 P_{l}=|\nu_{l}\rangle~\langle
 \nu_{l}|
\end{equation}
is the operator of the projection on  the flavor neutrino state
$|\nu_{l}\rangle $ ($\langle \nu_{l'}|\nu_{l}\rangle=\delta_{l'l}$, where
~$l=e,\mu,\tau$). It is obvious that
\begin{equation}\label{projection1}
 P_{l}^{2}=P_{l}~.
\end{equation}
The average value of the operator $P_{l}$ is given by
\begin{equation}\label{probability}
\overline{P_{l}}(t)=|\langle \nu_{l}|\Psi(t)\rangle|^{2}~.
\end{equation}
Thus, $\overline{P_{l}}(t)$ is the probability to find the flavor
neutrino $\nu_{l}$ in the state $|\Psi(t)\rangle$. We will assume
that $|\Psi(0)\rangle=| \nu_{l}\rangle$. In this case we have
\begin{equation}\label{survival}
\overline{P_{l}}(t)=P_{\nu_{l}\to \nu_{l}}(t)~,
\end{equation}
where $P_{\nu_{l}\to \nu_{l}}(t)$ is the probability of the flavor
neutrino $\nu_{l}$ to survive. Obviously we have
\begin{equation}\label{probability1}
P_{\nu_{l}\to \nu_{l}}(0)=1,\quad P_{\nu_{l}\to \nu_{l}}(t)\leq 1,
~~t>0~.
\end{equation}
Taking into account (\ref{projection1}), we have
\begin{equation}\label{deviation}
\Delta P_{l}(t) =\sqrt{P_{\nu_{l}\to \nu_{l}}(t)-P^{2}_{\nu_{l}\to
\nu_{l}}(t)}~.
\end{equation}
The inequality ({\ref{timeenergy1}) takes the form
\begin{equation}\label{timeenergy3}
\Delta E \geq \frac{1}{2}~\frac{|\frac{d }{d t}P_{\nu_{l}\to
\nu_{l}}(t) |}{\sqrt{P_{\nu_{l}\to \nu_{l}}(t)-P^{2}_{\nu_{l}\to
\nu_{l}}(t)}}~.
\end{equation}
We will consider the survival probability $P_{\nu_{l}\to
\nu_{l}}(t)$ in the interval
\begin{equation}\label{interval}
0\leq t\leq t_{\rm{1min}}~,
\end{equation}
where $t_{\rm{1min}}$ is the time at which the survival probability
reaches the first minimum. In this interval $\frac{d }{d
t}P_{\nu_{l}\to \nu_{l}}(t)\leq 0$. After the integration of the
inequality (\ref{timeenergy3}) over the time from $t=0$ to t we find
\begin{equation}\label{timeenergy4}
\Delta E~t \geq \frac{1}{2}~\left(\frac{\pi}{2}-\arcsin
(2~P_{\nu_{l}\to \nu_{l}}(t)-1)\right)~.
\end{equation}
The expressions (\ref{survival1}), (\ref{atmsurvival}) and
(\ref{twoneutrinoKL}) for the survival probabilities which, as we
stressed before, describe all existing experimental data, depend on the
distance $L$ between the neutrino production and detection points.
However, for ultrarelativistic neutrinos \footnote{The relation
(\ref{tequalL}) was confirmed by the K2K \cite{K2K} and MINOS
\cite{Minos} neutrino oscillation experiments. In the K2K experiment
neutrinos are produced in $1.1 ~\mu s $ spills. Protons are
extracted from the accelerator every 2.2 s. It was found that
$$-0.2 \leq \left(t-L/c\right)\leq 1.3~ \mu s~.$$
Here $t=(t_{SK}-t_{KEK})$, where $t_{KEK}$ is the time of the
neutrino production at the KEK accelerator and $t_{SK}$ is the time
of the neutrino detection in the SK detector. }
\begin{equation}\label{tequalL}
L\simeq t~.
\end{equation}
We will assume in accordance with (\ref{survival}) that the transition
probabilities depend on time. Let us consider $P_{\nu_{\mu}\to \nu_{\mu}}(t)$
for transitions in the atmospheric-LBL region which are
driven by $\Delta m_{23}^{2}$. From (\ref{atmsurvival}) and
(\ref{SKbest}) we have
\begin{equation}\label{minimumt1}
P_{\nu_{\mu}\to \nu_{\mu}}(t^{(23)}_{\rm{1min}})\simeq 0~,
\end{equation}
where
\begin{equation}\label{minimumt2}
t^{(23)}_{\rm{1min}}= 2\pi~\frac{E}{\Delta m_{23}^{2}}.
\end{equation}
From (\ref{timeenergy4}), (\ref{minimumt1}) and (\ref{minimumt2}) we
obtain the following time-energy uncertainty relation for the
neutrino oscillations driven by $\Delta m_{23}^{2}$
\begin{equation}\label{timeenergy5}
\Delta E~t_{\rm{osc}}^{(23)} \geq \pi~,
\end{equation}
where
\begin{equation}\label{oscperiod1}
t_{\rm{osc}}^{(23)}=2t^{(23)}_{\rm{1min}}=4\pi~\frac{E}{\Delta
m_{23}^{2}}
\end{equation}
is the period of neutrino oscillations in the atmospheric-LBL region.

The condition (\ref{timeenergy5}) is satisfied if the term
\begin{equation}\label{term}
 \frac{1}{2}~(1-\cos\Delta
m^2_{23} \frac {t} {2E})
\end{equation}
in (\ref{survival1}) is changed in the interval (\ref{interval})
from 0 to 1. In  the expression for the survival probability the term
(\ref{term}) is multiplied by the oscillation amplitude which is
determined by the mixing angle. In order for neutrino oscillations to be
observed not only the term (\ref{term}) has to be changed
significantly  in the interval (\ref{interval}) but also the mixing
angle must be relatively large. This means that if the mixing angle is
small the time-energy uncertainty relation (\ref{timeenergy5}) is
only a {\em necessary condition} for neutrino oscillations driven by
$\Delta m_{23}^{2}$ to be observed.

Driven by $\Delta m_{23}^{2}$, the probability of $\bar\nu_{e}$ to
survive in accordance with the results of the CHOOZ experiment is
close to one. Let us apply inequality (\ref{timeenergy4}) to
this case. From (\ref{baresurvival}) we have
\begin{equation}\label{choozmin}
P_{\bar\nu_{e}\to\bar\nu_{e}}(t^{(23)}_{\rm{1min}})=1-\sin^{2}2\theta_{13}~,
\end{equation}
where $t^{(23)}_{\rm{1min}}$ is given by (\ref{minimumt2}). From the
results of the CHOOZ experiment \cite{Chooz} follows that
\begin{equation}\label{choozbound}
\sin^{2}2\theta_{13}\lesssim 2\cdot10^{-1}
\end{equation}
Up to  $\sin^{3}2\theta_{13}$ terms we have  the following expansion
\begin{equation}\label{expansion}
\arcsin (2~P_{\bar\nu_{e}\to
\bar\nu_{e}}(t^{(23)}_{\rm{1min}})-1)\simeq\frac{\pi}{2}-2\sqrt{\sin^{2}2\theta_{13}}~.
\end{equation}
From (\ref{timeenergy4}) and (\ref{expansion}) we find for the
$\bar\nu_{e}\to\bar\nu_{e}$ transition the relation
\begin{equation}\label{timeenergy6}
\Delta E~t_{\rm{osc}}^{(23)} \geq 2~\sin2\theta_{13}~,
\end{equation}
which is much weaker than the time-energy uncertainty relation
(\ref{timeenergy5}).

Let us now consider $\bar\nu_{e}\to\bar\nu_{e}$  transitions in the
KamLAND region. These transitions are due to
$\bar\nu_{e}\leftrightarrows\bar\nu_{\mu,\tau}$ oscillations driven
by $\Delta m_{12}^{2}$. From (\ref{twoneutrinoKL}) we have
\begin{equation}\label{KLmin}
P_{\bar\nu_{e}\to\bar\nu_{e}}(t^{(12)}_{\rm{1min}})=1-\sin^{2}2\theta_{12}~,
\end{equation}
where
\begin{equation}\label{minimumt3}
t^{(12)}_{\rm{1min}}= 2\pi~\frac{E}{\Delta m_{12}^{2}}~.
\end{equation}
Further from (\ref{timeenergy4}) we find
\begin{equation}\label{timeenergy7}
\Delta E~t^{(12)}_{\rm{1min}} \geq
\frac{1}{2}~\left(\frac{\pi}{2}+\arcsin
(1-2~(1-\sin^{2}2\theta_{12}))\right).
\end{equation}
For the best-fit value of the parameter $\tan^{2}\theta_{12}$ we
have
\begin{equation}\label{bestfit}
\sin^{2}2\theta_{12}=0.87.
\end{equation}
With this value for $\sin^{2}2\theta_{12}$ we obtain from
(\ref{timeenergy7}) with an accuracy of $\sim 1\%$ the following
time-energy uncertainty relation in the KamLAND region
\begin{equation}\label{timeenergy8}
\Delta E~t^{(12)}_{\rm{osc}} \geq \left(\pi -2\sqrt
{(1-\sin^{2}2\theta_{12})}\right)~,
\end{equation}
where
\begin{equation}\label{oscperiod}
t^{(12)}_{\rm{osc}}=4\pi~\frac{E}{\Delta m_{12}^{2}}
\end{equation}
is the period of oscillations driven by $\Delta m_{12}^{2}$. Using
the value (\ref{bestfit}) we have
\begin{equation}\label{timeenergy9}
\Delta E~t^{(12)}_{\rm{osc}} \geq (\pi-0.72).
\end{equation}

\section{Recoilless creation and resonance absorption of  tritium
antineutrinos}

In \cite{Raghavan,Potzel} the possibilities have been considered to
perform an experiment on the detection of the tritium $\bar\nu_{e}$
with energy $\simeq$ 18.6 keV in the recoilless (M\"ossbauer)
transitions
\begin{equation}\label{mosstransitions}
 ^{3}\rm{H}\to ^{3}\rm{He}+\bar\nu_{e},\quad \bar\nu_{e}+
^{3}\rm{He}\to^{3}\rm{H}.
\end{equation}
It was estimated in \cite{Raghavan}  that the relative uncertainty
of the energy of the  antineutrinos produced in
(\ref{mosstransitions}) is of the order
\begin{equation}\label{energyuncert}
\frac{\Delta E }{E}\simeq 4.5 \cdot 10^{-16}~.
\end{equation}
With such an uncertainty it was estimated  \cite{Raghavan} that the
cross section of the recoilless resonance absorption of
antineutrinos in the process $\bar\nu_{e}+ ^{3}\rm{He}\to
^{3}\rm{H}$ is equal to
\begin{equation}\label{crosssection}
\sigma_{R}\simeq 3\cdot 10^{-33}\rm{cm}^{2}
\end{equation}
Such a value is about nine orders of magnitude larger than the
normal neutrino cross section.

For the tritium antineutrino with the energy $\simeq$ 18.6 keV the
length of the oscillations driven by $\Delta m^{2}_{23}$ is given by
\begin{equation}\label{osctime}
L^{(23)}_{osc}\simeq 2.5 \frac{E(\rm{MeV})}{\Delta
m^{2}_{23}(\rm{eV^{2})}}~m \simeq 18.6 ~m
\end{equation}
It was proposed in \cite{Raghavan} to search for neutrino
oscillations in a M\"ossbauer neutrino experiment. Such a
measurement would allow to determine the parameter
$\sin^{2}\theta_{13}$ (or to improve the CHOOZ bound (\ref{teta13})) in a
neutrino experiment with a baseline of about 10 m.

We will now discuss possibilities to observe neutrino oscillations in a
M\"ossbauer neutrino experiment from the point of view of the
time-energy uncertainty relations which we obtained in the previous
section.

From (\ref{timeenergy5}) follows that for neutrino oscillations,
which are driven by the "large" atmospheric $\Delta m^{2}_{23}$, the
energy uncertainty must satisfy the following condition
\begin{equation}\label{inequality}
\frac{\Delta E }{E}\gtrsim \frac{1}{4}~\frac{\Delta
m^{2}_{23}}{E^{2}}\simeq 1.8\cdot 10^{-12}~.
\end{equation}
Thus, the expected neutrino energy uncertainty (\ref{energyuncert})
in the M\"ossbauer neutrino experiment is about four orders of
magnitude {\em smaller} than the minimum energy uncertainty which is required
by the time-energy uncertainty relation (\ref{timeenergy5}) for
neutrino oscillations to occur.

It is of interest to see what energy uncertainty is required by the
uncertainty relation in its weak form for neutrino flavour oscillations in the atmospheric-LBL region. From (\ref{oscperiod1}) and (\ref{timeenergy6}) we
have
\begin{equation}\label{weakuncerainty}
\frac{\Delta E }{E}\gtrsim \frac{1}{2\pi}~\frac{\Delta
m^{2}_{23}}{E^{2}}\sin2\theta_{13}~.
\end{equation}
Using the CHOOZ
bound for the value of the parameter $\sin2\theta_{13}$ we have
\begin{equation}\label{weakuncerainty1}
\frac{\Delta E }{E}\gtrsim 0.5\cdot 10^{-12}~.
\end{equation}
In the future reactor and T2K experiments the sensitivity to the
parameter $\sin^{2}2\theta_{13}$ is planned to be improved by a
factor of 20 \cite{T2K}. If for the value of the parameter
$\sin^{2}2\theta_{13}$ we take the value
$\sin^{2}2\theta_{13}=10^{-2}$ we have from (\ref{weakuncerainty})
\begin{equation}\label{weakuncerainty2}
\frac{\Delta E }{E}\gtrsim 1.2\cdot 10^{-13}~.
\end{equation}
Thus, the neutrino energy uncertainty (\ref{energyuncert}) in a
M\"ossbauer neutrino experiment is several orders of magnitude
smaller than the minimum energy uncertainty which is required even by
the weak form of the time-energy uncertainty relation
(\ref{timeenergy6}).

We will now discuss neutrino oscillations in a M\"ossbauer
neutrino experiment which are driven by the "small" solar-KamLAND neutrino
mass-squared difference $\Delta m^{2}_{12}$ given by
(\ref{KLsolar}). In this case, the oscillation length for the tritium neutrinos
will be $L^{(12)}_{osc}\simeq 600 ~\rm{m}$. Thus, the
baseline of the experiment must be about 300-400 meters. This makes
such an experiment very difficult. Let us see, however, whether the
experiment is possible from the point of view of the time-energy
uncertainty relation. From (\ref{timeenergy9}) we have
\begin{equation}\label{inequality1}
\frac{\Delta E }{E}\gtrsim \frac{(\pi-0.72)}{4\pi}~\frac{\Delta
m^{2}_{12}}{E^{2}}\simeq 4.2\cdot 10^{-14}~,
\end{equation}
Thus, the neutrino energy uncertainty required by the time-energy
uncertainty relation for neutrino oscillations driven by $\Delta
m^{2}_{12}$ must also be larger than the estimated energy uncertainty in
the M\"ossbauer neutrino experiment.

We compared the required minimal neutrino energy uncertainty
$\frac{\Delta E }{E}$ with the value (\ref{energyuncert}) estimated
in \cite{Raghavan}. However, it was stressed  in \cite{Potzel} that
due to inhomogeneous broadening (impurities, lattice defects and
other effects), the real value for $\frac{\Delta E }{E}$ can
be still larger than that in (\ref{energyuncert}) and inequality (\ref{inequality1}) could be
satisfied. However, the maximal resonance effect would then be
reduced accordingly. Because of the large
baseline, such a M\"ossbauer neutrino experiment on the
investigation of neutrino oscillations driven by the mass-squared
difference $\Delta m^{2}_{12}$ does not look feasible.

\section{Conclusion}
In spite of neutrino oscillations having been observed in atmospheric, solar,
reactor and accelerator neutrino experiments the nature of neutrino
oscillations is still an open problem. The M\"ossbauer neutrino
experiment proposed in \cite{Raghavan,Potzel} gives a unique
possibility to test the origin of neutrino oscillations
\cite{BilFeilPotz07}.

We consider here M\"ossbauer neutrino experiments from the point of
view of the time-energy uncertainty relation, which connects the
uncertainty of the neutrino energy with a characteristic time of
neutrino oscillations. Using the general Mandelstam-Tamm method,
which is based only on the assumption that the evolution of Heisenberg
operators in quantum field theory is determined by the Hamiltonian,
we derived time-energy uncertainty relations for neutrino
oscillations. We conclude that the small energy uncertainty of
M\"ossbauer antineutrinos recoillessly emitted and absorbed in the
$^{3}\rm{H}/^{3}\rm{He}$ system is in conflict with the energy
uncertainty which, according to the time-energy uncertainty
relation, is necessary for neutrino oscillations to happen.

There exist other approaches to neutrino oscillations which do allow
oscillations in the M\"ossbauer neutrino experiment (effectively due
to phase differences of the states of neutrinos with different
masses, same energy and different momenta; see the recent paper
\cite{Akhmedov}, in which the propagation of virtual mixed neutrinos
between the source and the detector was considered in detail, and also
papers \cite{Stodol,Lipkin,Kayser})).

In paper \cite{TammMand45} it is stated: "From definiteness
of the total energy of a system follows the constancy in time of all
dynamical variables." If practically monochromatic M\"ossbauer
neutrinos oscillate this would mean that neutrino oscillations do
not follow this general quantum rule. Oscillations of M\"ossbauer
neutrinos would indicate a stationary phenomenon where the evolution
of the neutrino state proceeds in space rather than in time. Thus,
the search for neutrino oscillations in the M\"ossbauer neutrino
experiment gives us a unique possibility to test fundamentally
different approaches to neutrino oscillations. Let us stress that
such tests can not be realized in usual neutrino oscillation
experiments (see \cite{BilFeilPotz07}).

In conclusion, we will summarize our main assumptions:
\begin{itemize}
\item We consider neutrino oscillations in the framework of QFT.
  \item We assume that in weak processes, flavor neutrinos are
  produced and detected in mixed mass eigenstates. The states of flavor neutrinos are
  given by vectors (\ref{flavorstates}).
  \item The Mandelstam-Tamm method, which we use, is based on the assumption that the dynamics of a quantum system
  is determined by the Schr\"odinger equation.
  \item We assume that for ultra-relativistic neutrinos, the average
  difference between the emission and absorption times is given by the
 distance between source and detector.
\end{itemize}
S.B. acknowledges the ILIAS program for support and the TRIUMF
theory department for the hospitality. This work has been supported by funds of the DFG (Transregio 27: Neutrinos and Beyond), the Munich Cluster of Excellence (Origin and Structure of the Universe), and the Maier-Leibnitz-Laboratorium (Garching).

\end{document}